\documentclass[%
 aip,
 amsmath,amssymb,
 reprint,%
]{revtex4-1}

\usepackage{graphicx}
\usepackage{dcolumn}
\usepackage{bm}

\usepackage[utf8]{inputenc}
\usepackage[T1]{fontenc}
\usepackage{mathptmx}
\usepackage{color}

\newcommand{\BXf}{B_{\mbox{\tiny Xf}}}
\newcommand{\muPol}{\mu_{\mbox{\tiny Pol}}}

\newcommand{\muEu}{\mu_{\mbox{\tiny Eu}}}
\newcommand{\JXf}{J_{\mbox{\it{\scriptsize Xf}}}\,}
\newcommand{\JXfS}{J_{\mbox{\it{\scriptsize Xf}}}S\,}
\usepackage{float}
\restylefloat{table}

\begin{document}

\preprint{AIP/123-QED}

\title{Hypergiant spin polarons photogenerated in ferromagnetic europium chalcogenides}

\author{X. Gratens}
\affiliation{Instituto de Física, Universidade de São Paulo, 05508-090 São Paulo, Brazil}
\author{Yunbo Ou}
\affiliation{Francis Bitter Magnet Lab and Plasma Science and Fusion Center, Massachusetts Institute of Technology, 77 Mass Ave., Cambridge, MA 02139,USA}
\author{J. Moodera}
\affiliation{Francis Bitter Magnet Lab and Plasma Science and Fusion Center, Massachusetts Institute of Technology, 77 Mass Ave., Cambridge, MA 02139,USA}
\affiliation{Department of Physics, Massachusetts Institute of Technology, 77 Mass Ave., Cambridge, MA 02139, USA}
\author{P. H. O. Rappl}
\affiliation{LAS-INPE, Av. dos Astronautas, 1758-Jd. Granja, 12227-010 São José dos Campos, Brazil}
\author{A. B. Henriques}
\email[]{andreh@if.usp.br}
\affiliation{Instituto de Física, Universidade de São Paulo, 05508-090 São Paulo, Brazil}

\date{\today}

\begin{abstract}

We find that in the ferromagnetic semiconductor EuS, near its Curie temperature, a single band-edge photon generates a spin polaron (SP), whose magnetic moment approaches 20,000 Bohr magnetons. This is much larger than the supergiant photoinduced SPs in antiferromagnetic europium chalcogenides, reported previously. The larger SP in ferromagnetic EuS, and still larger expected for EuO, is explained by a larger Bohr radius of the photoexcited electron state, which encircles and polarizes a greater number of lattice spins. However, because the wave function of the photoexcited electron spreads over a greater volume, the photoexcited electron’s exchange interaction with individual lattice spins weakens, which makes the SP more easily quenched thermally.
\end{abstract}
\maketitle

\begin{quotation}
Uncovering new efficient techniques by which light magnetizes materials is an interesting scientific modern topic, and paves the way for novel optomagnetic devices \cite{kirilyukRMP, nature2019}. Recently, an ultrafast magnetization mechanism based on the photoinduction of spin polarons for intrinsic antiferromagnetic europium chalcogenides EuTe and EuSe was reported \cite{apl2011,prb2014,prb16Rapid,prb17EuTe,prb17FR,prl2018}.
The magnetization process is described as follows. A photon resonant with the bandgap generates a randomly oriented spin polaron (SP). The magnetic moment of the photoinduced SP is of the order of $600$ Bohr magnetons in EuTe and $6,000$ in EuSe. These photoinduced SPs are orders of magnitude larger than those seen in
diluted magnetic semiconductors \cite{wolff,prlPolaronSize} or detected by
muon investigations \cite{storchakmuons}, which are of a few tens
of Bohr magnetons only. Due to the large magnetic moment of the SPs in EuTe and EuSe, a very small magnetic field is sufficient to align their spin with the field, magnetizing the sample. The remarkable characteristic of this process is that the absorption of a single photon leads to the spin alignment of thousands of electrons, thus opening the perspective of controlling magnetism with very low intensity light.
\end{quotation}

In this paper, we demonstrate that in the ferromagnetic members of the europium chalcogenide series, a single photon generates an even larger SP than in the antiferromagnetic ones, inducing a spin coherence of tens of thousands of Bohr magnetons.
The much larger intrinsic SP in the ferromagnetic members, EuO and EuS, than in the antiferromagnetic ones, EuSe and EuTe, is not entirely surprising. The SP in europium chalcogenides is in essence an effective mass exciton, within which the band-lattice exchange interaction enhances ferromagnetic spin arrangement. The hole belongs to the strongly localized $4f$ shell of an europium atom. In contrast to the hole, the electron in the exciton extends over many lattice parameters, and is described by a Bohr envelope wave function, with an effective Bohr radius
\begin{equation}
a_B=\frac{\varepsilon\hbar^2}{m^\ast ke^2},
\label{eq:aB}
\end{equation}
where $\varepsilon$ is the dielectric constant of the host material, $m^\ast$ is the effective mass of the electron in the conduction band, $k=9.0\times 10^9$N-m$^2$/C$^2$ is Coulomb's constant, and $e=1.60\times 10^{-19}$~C is the elementary charge. The Coulomb attraction by the hole is crucial to stabilize the SP \cite{maugerTB,mauger,prb2014}. However, the strong intra-atomic localization of the hole suppresses its exchange interaction with surrounding lattice spins, and it does not participate in the lattice spin polarization. The strong localization of the hole also makes the SP immobile.

We can roughly estimate the potential size of photoinduced SPs in EuX by assuming full ferromagnetic alignment within the exciton’s Bohr sphere, which gives a magnetic moment of
\begin{equation}
\muPol\sim \frac{4}{3}\pi a_B^3 \,\,N\muEu,
\label{eq:mupol}
\end{equation}
where $\muEu=7\mu_B$ is the magnetic moment of an Eu atom, $\mu_B=9.27\times 10^{-24}$~A.m$^2$ is Bohr's magneton, and $N=\frac{4}{a^3}$ is the density of Eu atoms in the EuX face-centered cubic lattice of parameter $a$.

\begin{table*}[!ht]
\caption{\label{tab:table1}
Predicted order-of-magnitude of photoinduced spin polarons, using equation~\eqref{eq:mupol}. Measured SP magnetic moments are also shown. Numbers in brackets indicate source references.}
\begin{ruledtabular}
\begin{tabular}{ccccccc}
EuX & $\varepsilon$ & $a_B$ (\AA) & $a$(\AA) & $\muPol/\mu_B$ & $\muPol/\mu_B$ & $\BXf (r=a_B)$ (T)\\
  &   & from eq. \eqref{eq:aB} & from Ref.[\onlinecite{mauger}]  & from eq. \eqref{eq:mupol} & Measured & from eq. \eqref{eq:bxf} \\
\hline
EuTe & 6.9 [\onlinecite{axe1060}] & 12.2 & 6.598 & 730      & 610  [\onlinecite{apl2011,prb2014,prb16Rapid,prb17EuTe}] & 1.2\\
EuSe & 9.4 [\onlinecite{axe1060}] & 16.6 & 6.195 & 2,300    & 6000 [\onlinecite{prl2018}] & 0.4\\
EuS & 11.1 [\onlinecite{axe1060}]  & 19.6 & 5.968 & 4,100  & 20,000 [This work] & 0.22\\
EuO & 26 [\onlinecite{guntherodt74}]   & 45.8 & 5.141 & 83,000 &  not yet known & 0.01\\
\end{tabular}
\end{ruledtabular}
\end{table*}

The effective mass was taken to be $m^\ast=0.3$, in units of the free electron mass, $m_0=9.1\times 10^{-31}$~kg, for all members of the EuX family \cite{cho,umehara99}. Table~\ref{tab:table1} shows the remaining parameters entering equations \eqref{eq:aB} and \eqref{eq:mupol}, and the estimated photoinduced SP produced by equation \eqref{eq:mupol}. 
Also shown in table~\ref{tab:table1} is the measured maximum size of photoinduced SPs. Table~\ref{tab:table1} shows that, for ferromagnetic EuO, equation \eqref{eq:mupol} predicts a SP approaching 100,000 Bohr magnetons. This hypergiant size is a consequence of the increasing ratio between the effective Bohr radius and the lattice constant. The greater the ratio, the greater the number of Eu spins within the Bohr sphere that can be potentially polarized to form a SP.


To test our predictions, photoinduced SPs were investigated in an epitaxial EuS sample, grown on a BaF$_2$ (111) substrate.
The BaF$_2$ substrate was cleaved of from a bar along (111) plane in air, and cleaned by dry nitrogen gas before being loaded into an UHV chamber. The substrate was annealed at 800$^\circ$C for 30 minutes to clean the surface, and then kept at 600$^\circ$C. EuS (99.9\%) was deposited at 600$^\circ$C by a commercial electron beam evaporator at the rate of 0.1 Å/s. After deposition, the film was annealed at 600$^\circ$C for 30 minutes. A 5~nm Al$_2$O$_3$ cap was deposited onto the hygroscopic EuS film, to protect it from contact with air.

\begin{figure}
\includegraphics[angle=0,width=86mm]{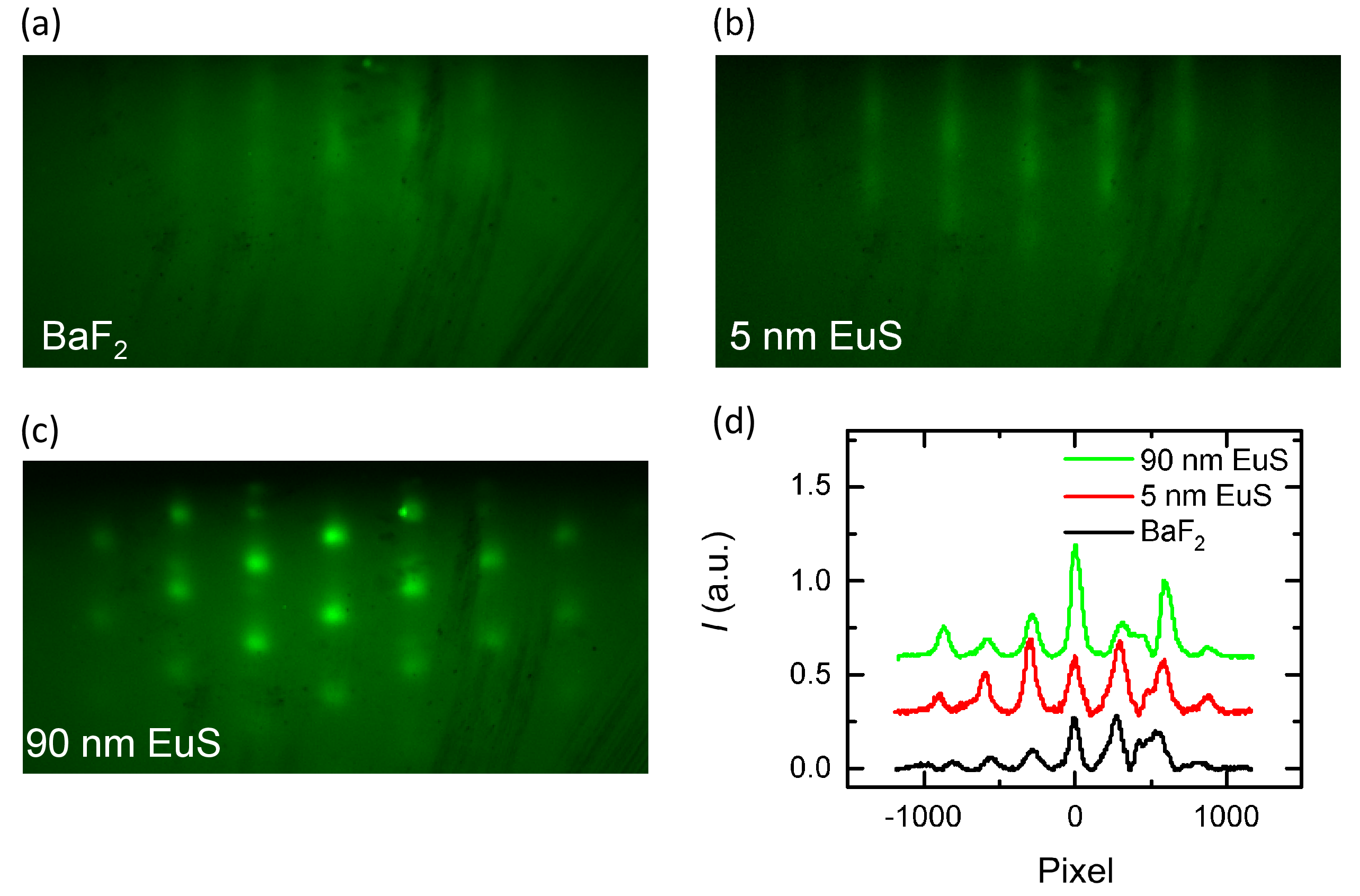}
\caption{{\em In situ} RHEED pattern of (a) BaF2 (111), (b) 5 nm EuS and (c) 90 nm EuS along $\Gamma-$K. (d) Line profile of (a)-(c). 
}
\label{fig:RHEED} 
\end{figure}

Because of the lattice mismatch of less than 4\%, (lattice constant $a$=5.968~\AA\ and
$a$=6.196\AA\ for EuS and BaF$_2$, respectively), the epitaxial layers are expected to have a high structural quality. 
Our {\it in situ} Reflective High-Energy Electron Diffraction (RHEED) confirmed the epitaxial growth of EuS on BaF$_2$ (111) substrate (see Figure~\ref{fig:RHEED}). 
The sharp streaky RHEED pattern for the 5 nm thick EuS film, shown in Fig.~\ref{fig:RHEED}(b), indicates a 2D growth. After 10 nm, due to the relaxation of strain, growth changes to a 3D mode, as revealed by the spotty RHEED pattern shown in Fig.~\ref{fig:RHEED}(c).
A 60$^\circ$ rotation symmetry confirmed by RHEED indicates EuS film growth along $\left< 111 \right>$ direction.

Next, we address the fundamental optical and magnetic properties of the EuS sample. 
All members of the EuX series have a common band-edge electronic energy level structure \cite{wachter,mauger}, independent of their natural magnetic arrangement. 
The primary effect of a change in magnetic order on the band-edge energy level structure is a rigid shift of conduction band energy levels relative to the top valence state \cite{prb08,ijmp2009}, which is a consequence of the ferromagnetic electron-lattice exchange interaction. 
Thus, for example, on cooling through the Curie temperature \cite{wachter}, the band gap of ferromagnetic EuO shrinks due to the onset of ferromagnetic order. Similarly, when a magnetic field imposes ferromagnetic order, the band gap of antiferromagnetic EuX shrinks  \cite{jpc04,jpc07,prb08}. 

Circular dichroism in EuX reflects the magnetic moment of the sample in the direction of light travel \cite{jpc08}. In zero magnetic field, ferromagnetic EuX naturally forms a multiple microdomain structure, and no net magnetic moment exists. In antiferromagnetic EuX, in zero magnetic field, the magnetic moment is always zero. Therefore circular dichroism in zero field is absent for all EuX members. However, a strong enough magnetic field turns EuX into a single ferromagnetic domain, and a huge magnetic circular dichroism develops \cite{prb05,jpc07,ijmp2009,prb08}.

\begin{figure}
\includegraphics[angle=0,width=86mm]{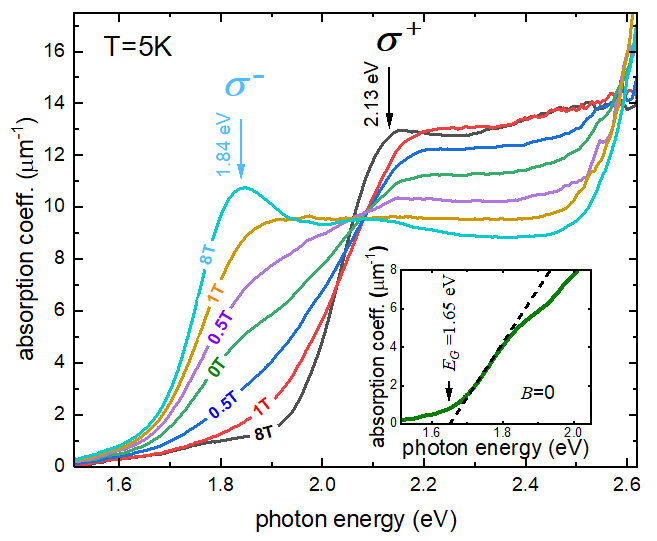}
\caption{Magnetic circular dichroism of the EuS sample at $T=5$~K. The inset shows the band-edge at $B=0$. 
}
\label{fig:MCD} 
\end{figure}

To verify if our EuS sample displayed the expected circular dichroism, we measured its band-edge polarized optical absorption at $T=5$~K, and the result is shown in figure~\ref{fig:MCD}.
Since $T=5$~K is well below $T_C$ for EuS (see below), the sample is in the ferromagnetic phase. Despite ferromagnetic order, circular dichroism is absent for $B=0$, because the photon crosses many magnetic domains, oriented randomly. However, when a magnetic field imposes a single domain oriented along the light wave vector, the huge circular dichroism inherent to EuX emerges, demonstrating the excellent quality of the sample. At saturation, the absorption onset shows a $\sigma^-$ peak at 1.84~eV, and a $\sigma^+$ peak at 2.13~eV. The size of the $\sigma^+-\sigma^-$ splitting is nearly the same as reported for other EuX members \cite{prb05,prb08}. From the spectrum at $B=0$, the EuS bandgap is estimated to be 1.65~eV (see inset of figure ~\ref{fig:MCD}). For europium chalcogenides, the band gap 
is found by extrapolation of the linear absorption region to zero. The EuX band gap determined in this way is very consistent, and circumvents shortcomings of Tauc plots or other approaches, as discussed in detail in Ref.~\onlinecite{jap2019}.
The changes in the shape of the optical spectrum with magnetic field is a consequence of the changes in the oscillator strengths of band-edge electronic transitions, due to the alignment of spin domains with the light wave vector \cite{prb05,prb08}. Nevertheless, in the case of EuS at $T=5$~K, the energy band gap itself does not change with $B$, because at this temperature the arrangement is always ferromagnetic.

\begin{figure}
\includegraphics[angle=0,width=86mm]{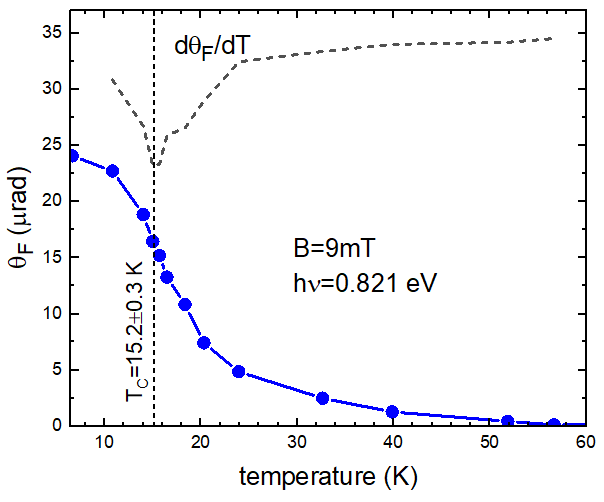}
\caption{Faraday rotation by the bulk of sample, at $B=0.009$~T, for a photon energy of 0.821~eV. The dotted line is the derivative, $d\theta_F/dT$, whose minimum yields the Curie temperature.}
\label{fig:FvsT} 
\end{figure}

In europium chalcogenides, in any magnetic phase, the magnetization of the sample is directly proportional to the bulk Faraday rotation angle, $\theta_F$, of light crossing the sample, if the wavelength is within the band gap \cite{jap2019}.
The proportionality constant is almost independent of temperature; therefore, the magnetization dependence on $T$ can be inferred from $\theta_F$ measurements. Figure~\ref{fig:FvsT} shows the $\theta_F$ vs $T$ trace obtained for $B$ = 9 mT, using a photon energy of 0.821~eV, which is well within the bandgap of 1.65~eV (see figure~1). Measurements started at room temperature and the sample was slowly cooled. The magnetization of the sample increased as the sample went through the Curie temperature and approached saturation, as expected for a ferromagnet, due to the appearance of spontaneous magnetization. The weak magnetic field of 9~mT was required to inhibit multiple domains that would preclude a macroscopic measurement.
The Curie temperature for our sample was taken to be equal to the temperature at which the derivative $d\theta_F/dT$ is minimum, giving $T_C = 15.2 \pm 0.3 $ K. 
The $T_C$ value for our thin epitaxial layer is slightly less than the known $T_C=16.3$~K for bulk EuS \cite{wachterEuS}.

We investigated photoinduced spin polarons in the EuS sample, using a two-color pump-probe Faraday rotation experiment \cite{prb16Rapid}. Electron-hole pairs were generated by photons above the bandgap (2.33~eV), and the photoinduced Faraday rotation (PFR) was measured for photons within the bandgap (1.46~eV). For $T>T_C$, the lattice spins are uncorrelated, hence the Weiss field is absent. In this case, photoinduced SPs rotate freely. We used low pump intensities, so that the average distance between adjacent SPs is much larger than their radius, making them non-interacting. Therefore, the SP ensemble forms a paramagnetic gas, whose magnetization is described by a Langevin function \cite{prb16Rapid,prb17EuTe}.

\begin{figure}
\includegraphics[angle=0,width=86mm]{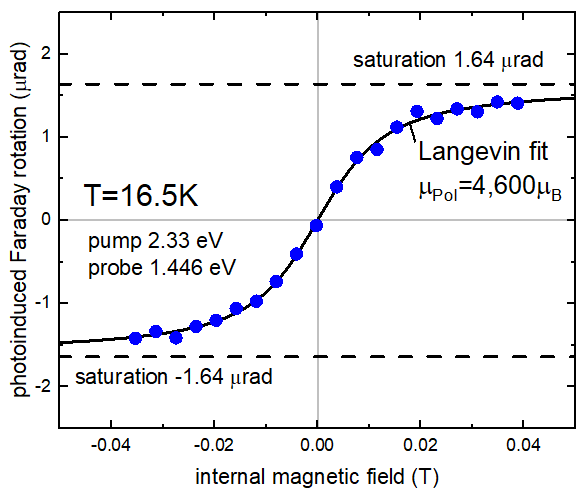}
\caption{Photoinduced Faraday rotation in EuS just above the Curie temperature (dots). The full line is a fit  with a Langevin function, which yields the amplitude and the SP magnetic moment.}
\label{fig:PFR} 
\end{figure}

Inevitably, the pump pulse causes heating of the illuminated, which will change the magnetization, hence it will show as PFR. Ref.~\onlinecite{prb17EuTe} reports a detailed investigation of the effect of pump heating on the PFR signal: the heating adds a PFR signal whose absolute value increases linearly with $B$.
Such a linear component is easily distinguished from the step-like photoinduced magnetization associated with photoinduced SPs \cite{prb17EuTe}. 
The dots in figure \ref{fig:PFR} represent the net PFR signal, associated with photoinduced SPs at 16.5~K, as a function of internal magnetic field. The solid line in figure \ref{fig:PFR} shows a fit of the net PFR signal with a Langevin function. The excellent fit yields $\muPol=4,600\,\mu_B$. The Langevin fit also yields a PFR angle at saturation of $1.64\,\mu$rad, as shown in figure \ref{fig:PFR}.

As discussed above, the Faraday rotation angle is proportional to the magnetization \cite{jap2019}. On the other hand, the magnetization of the polaron ensemble, at saturation, is proportional to the magnetic moment of an individual SP. 
Therefore, from figure~\ref{fig:PFR}, the factor required to convert saturation PFR into SP magnetic moment is 4,600/1.64=2,800~$\mu_B/\mu$rad. 
This factor can be used to convert the measured saturation PFR into the SP magnetic moment at all temperatures, provided that all measurements be taken for the same pump intensity. This
includes $T<T_C$, when SP orientation may be constrained by the onset of the ferromagnetic phase. Below $T_C$, a spontaneous magnetization arises, which gives rise to a Weiss field, which may constrain the rotation freedom of the SPs, 
hindering Langevin magnetism. 
\begin{figure}
\includegraphics[angle=0,width=86mm]{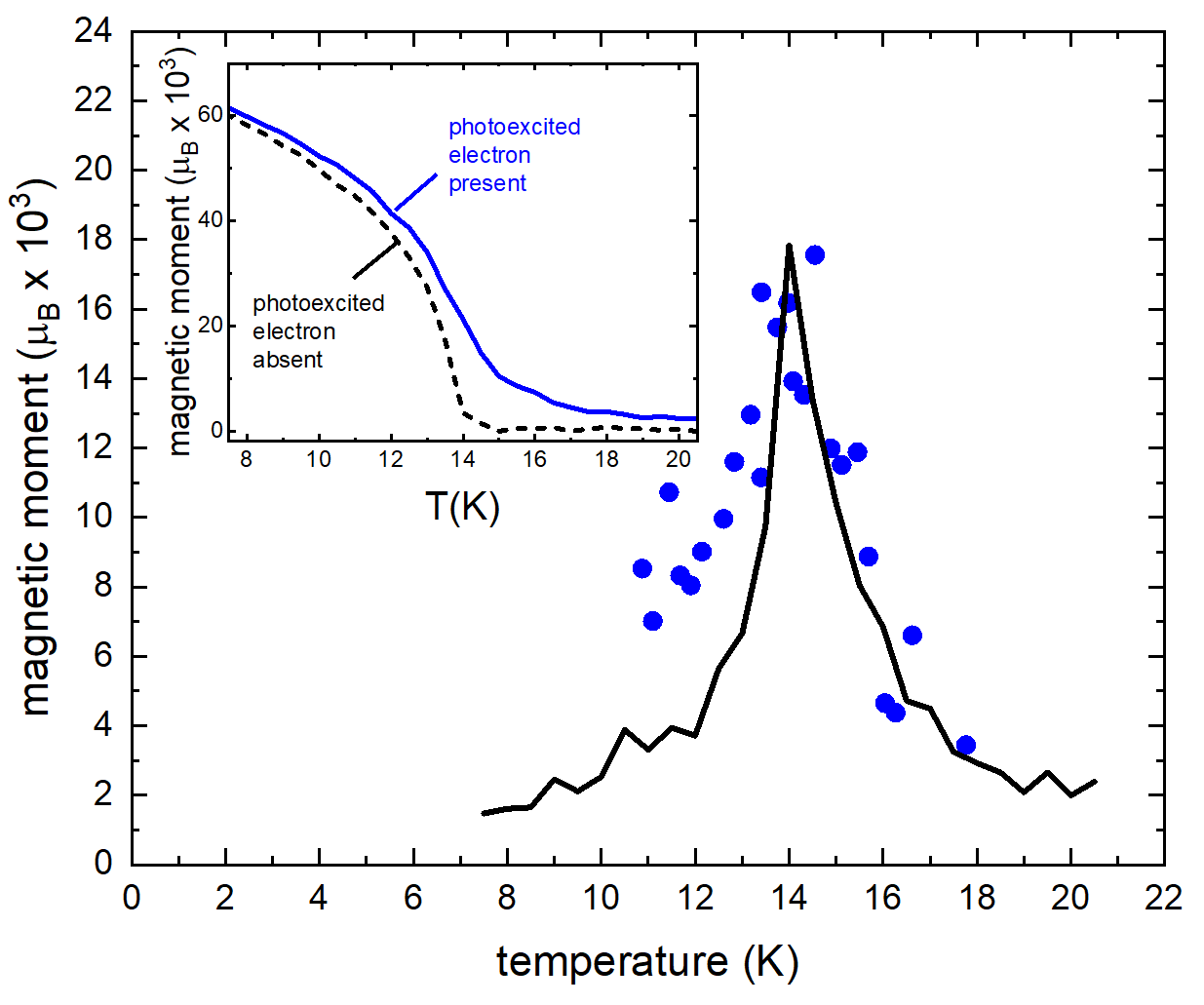}
\caption{The inset shows the magnetic moment of a spherical region, with and without a photoexcited electron, calculated by the Monte Carlo method. The difference between the two is depicted by the full curve, representing the theoretical magnetic moment of a SP. Dots depict the measured magnetic moment of the photoinduced SP as a function of temperature. }
\label{fig:muPolVsT} 
\end{figure}

Figure \ref{fig:muPolVsT} shows the measured magnetic moment of the photoinduced SP in EuS, as a function of temperature. Above $T_C$, the magnetic moment of the SP was determined from a Langevin fit to the PFR, whereas for $T<T_C$ the SP magnetic moment was obtained from the saturation PFR, using the conversion factor of 2,800~$\mu_B/\mu$rad described above. 

The photogenerated SP size shows a sharp peak near the Curie temperature.
This contrasts to antiferromagnetic EuTe, where the SP maximum size is two orders of magnitude smaller, and its size varies very slowly with temperature \cite{prb16Rapid}. This enormous difference in behavior is explained by the greater effective Bohr radius in EuS than in EuTe (table~\ref{tab:table1}), as follows. The raison d'\^etre of SPs is the exchange interaction between the photoexcited electron and the surrounding Eu spins. The exchange interaction is described by an effective magnetic field, $\BXf(r)$, acting on the lattice spins \cite{prb2014}
\begin{equation}
\BXf(r)=\frac{\JXf S}{N\muEu}|\psi(r)|^2,    
\label{eq:bxf}
\end{equation}
where $\JXf$ is the band-lattice exchange interaction constant \cite{apl2011},  $\Psi(r)=\frac{e^{-r/a_B}}{\sqrt{\pi a_B^3}}$ is the Bohr envelope wave function describing the photoexcited electron, and $r$ is the distance to the center of the SP. In EuTe, the average effective
magnetic field within the SP is about 1~T \cite{apl2011,prb2014}, which is approximately equal to $\BXf(a_B)$ (see last column in table~\ref{tab:table1}). When $a_B$ increases, on the one hand, the electron wave function covers an increasing number of spins, producing a larger SP, but on the other hand, the average effective magnetic field acting on an individual lattice spin drops, which reduces the polarization power of the photoexcited electron. As table~\ref{tab:table1} shows, in EuS, the typical exchange magnetic field inside the SP is only about 0.2~T. 
Nevertheless, the reduced average exchange magnetic field is still enough to polarize the lattice spins in the close vicinity to $T_C$, when the magnetic susceptibility is very high \cite{blundellbook}.  To either side of the Curie temperature, the magnetic susceptibility decreases, and the magnetic moment of an SP decreases in size. Well below $T_C$, the spontaneous magnetization saturates, which completely quenches SP formation.

To provide support for our interpretations, we performed Monte Carlo calculations. 
A cube of 18 EuS lattice parameters, centered on an Eu atom, was analyzed. A randomly oriented $S$=7/2 spin was associated with all Eu sites in the cube. The exchange interaction between lattice spins was incorporated via the first and second neighbor exchange constants, $J_1$ and $J_2$, respectively.
To exclude surface effects, Born-Karm\'an periodic boundary conditions were imposed on the cube. The exchange interaction between Eu spins and the photoexcited electron was incorporated via the effective magnetic field acting on lattice spins, $\BXf(r)$, given by eq.~\eqref{eq:bxf}. Then every lattice spin in the cube was reoriented individually, using the Monte Carlo stochastic method; this process was repeated 1,600 times at each temperature. The magnetic moment of the SP was taken to be the difference between two magnetic moment calculations of a sphere centered on the cube, one including and the other excluding the exchange field produced by the photoexcited electron (see inset in figure~\ref{fig:muPolVsT}). The radius of the sphere was determined from the condition $\BXf(r)>B_0$. Here, $B_0$ is the minimum value of the exchange field, produced by the photoexcited electron, required to overpower other competing interactions \cite{prb2014}.
One obvious competing field is the applied one. As figure~\ref{fig:PFR} shows, measurements of PFR requires the application of fields equivalent to an internal field of a few tens of milliteslas. In the region where the exchange field is
much less than the applied field, the photoexcited electron loses the competition and does not polarize the lattice spins. We assumed $B_0=0.005$~T. The Monte Carlo result 
is depicted in figure~\ref{fig:muPolVsT} by the full curve, using $J_1=0.2$~K,  $J_2=-0.09$~K, and $\JXfS=0.3$~eV. The calculated SP magnetic moment curve shows a peak with both maximum value and position in excellent agreement with the experiment. The parameters $J_1$, $J_2$ are $\sim$10\% less than values reported in the literature for bulk EuS \cite{euSJ1J2,prb2014}, the small difference could be associated with residual strain in our epitaxial sample \cite{mirabeau}.

Next, we shall compare the measured SP size at $T_C$ to the value expected by the molecular field (Weiss field) approximation \cite{reifBook}. In this model, the magnetic moment direction of each atom in the crystal is replaced by an average over the ensemble. Then, the magnetization  as a function of temperature and field is given by the transcendental equation \cite{blundellbook}
\begin{equation}
M(B,T)=N\mu B_S\left[\frac{\mu (B+\lambda M)}{k_BT}\right],
\label{eq:transc}
\end{equation}
where $B_S$ represents a Brillouin function of order $S$, $\mu$ is the absolute value of the magnetic moment of the atoms in the lattice, and $k_B=1.38\times 10^{-23}$~J/K is the Boltzmann constant. The term $\lambda M$ in equation~\eqref{eq:transc}
represents the molecular field, which describes the ferromagnetic coupling between spins. However, the molecular field model neglects correlated fluctuations in the orientation of 
different spins. Despite this drastic approximation, the molecular field model remarkably 
exhibits all the main features of ferromagnetism, including the temperature dependence of the magnetization on going through a paramagnetic/ferromagnetic phase transition in Nickel for example \cite{reifBook}, as discussed in Ref.~\onlinecite{reifBook} (p.~433-434). The molecular field model also describes the magnetization dependence on magnetic field at the Curie temperature in ferromagnets (see, for instance, Ref.~\onlinecite{blundellbook}, page~90), which is exactly the approach we shall use here.

Expanding the Brillouin function in a power series, $B_S(x)=\alpha_S x - \beta_S x^3$, where $\alpha_S=\frac{S+1}{3S}$ and $\beta_S=\frac{(S+1)^2+S^2}{90S^3}(S+1)$,
and substituting $\lambda=\frac{k_BT_C}{\alpha_SN\mu^2}$ \cite{blundellbook}, equation \eqref{eq:transc}
can be resolved for the magnetization at $T=T_C$, at a field $B$:
\begin{equation}
M(B,T_C)=N\mu\left(\frac{\alpha_S^4\mu\,B}{\beta_S k_BT_C}\right) ^{1/3}. 
\label{eq:MTCofB}
\end{equation}

Substituting $B$ in \eqref{eq:MTCofB} by $\BXf(r)$, as given by equation \eqref{eq:bxf}, and integrating \eqref{eq:MTCofB} in a sphere where $\BXf(r)>B_0$, we obtain $\muPol$ at $T=T_C$, in units of the Bohr magneton:
\begin{equation}
\muPol=
27\,A\left(\frac{16\pi^2\alpha_S^4}{\beta_S}\right)^{1/3}
\left(\frac{a_B}{a}\right)^2\left(\frac{\JXfS}{k_BT_C}\right)^{1/3}\,\mu,
\label{eq:muPol_Weiss}
\end{equation}
where $A=1-\left(\frac{x_0^2}{2}+x_0+1\right)e^{-x_0}$, and $x_0=\frac{1}{3}\ln\frac{\JXfS}{28\pi\mu_B B_0}\left(\frac{a}{a_B}\right)^3$.

To estimate the SP size at $T_C$ for EuS using (6), we substitute
$S=7/2$, $\mu=\muEu$, $a_B/a=3.28$ from table~\ref{tab:table1}, the experimental value $T_C=15.2$K, $\JXfS=0.30$~eV, and $B_0=0.005$~T, as used in the Monte Carlo calculations, to obtain $\muPol=20,000\,\mu_B$. This value agrees well the estimates given in Figure~\ref{fig:muPolVsT}. 
Of course, because it is based on the exclusion of fluctuations, equation \eqref{eq:muPol_Weiss} may not be accurate, nevertheless if provides a very useful simple practical formula for a first estimate of what kind of SP size can be expected in any concentrated ferromagnetic semiconductor, if its basic parameters are known.

In conclusion, we demonstrated that in the ferromagnetic members of the europium chalcogenide series,  
the magnetic moment of a photoinduced spin polaron can be much larger than the supergiant spin polaron observed in the antiferromagnetic ones. However, the hypergiant SP photogeneration in ferromagnetic EuX is only efficient in a narrow temperature interval around the Curie temperature. 
Above $T_C$, thermal fluctuations rapidly destroy the SPs. Below the critical phase transition temperature, in the ferromagnetic members, the onset of spontaneous magnetization quenches SP formation, whereas in the antiferromagnetic ones, SPs are formed most efficiently. For EuO, the exciton effective Bohr radius is larger than in EuS, so potentially an even larger SP is possible in EuO. However, because the Curie temperature for EuO ($T_C$=69~K) is higher, thermal quenching could be more effective in EuO than in EuS. This remains to be clarified experimentally.

\section{Acknowledgments}
This work was funded by
CNPq (Projects 303757/2018-3 and 420531/2018-1),
FAPESP (Projects 2019/02407-7 and 2019/12678-8), the National Science Foundation (NSF-DMR 1700137), Office of Naval Research (N00014-16-1-2657), and Center for Integrated Quantum Materials (DMR-1231319).

\bibliography{2020APLEuS}

\end{document}